\documentclass{article}
\usepackage{multirow}

\usepackage{arxiv}
\usepackage{amsmath,amsfonts}
\usepackage[utf8]{inputenc} 
\usepackage[T1]{fontenc}    
\usepackage{hyperref}       
\usepackage{url}            
\usepackage{booktabs}       
\usepackage{amsfonts}       
\usepackage{nicefrac}       
\usepackage{microtype}      
\usepackage{lipsum}
\usepackage{graphicx}
\graphicspath{ {./images/} }

\title{Physics-Driven Neural Compensation For Electrical Impedance Tomography}

\author{
 Chuyu Wang \\
  School of  Biomedical engineering\\
  University of Science and Technology of China\\
  Suzhou, 215123, Jiangsu, China\\
  \texttt{chuyuwang@mail.ustc.edu.cn} \\
   \And
 Huiting Deng \\
 School of  Biomedical engineering\\
  University of Science and Technology of China\\
  Suzhou, 215123, Jiangsu, China\\
  huitingdeng@mail.ustc.edu.cn
  \And
 Dong Liu \\
 CAS Key Laboratory of Microscale Magnetic Resonance\\
  University of Science and Technology of China\\
  Hefei 230026, China\\
  dong.liu@outlook.com
}

\begin{document}
\maketitle
\begin{abstract}
Electrical Impedance Tomography (EIT) provides a non-invasive, portable imaging modality with significant potential in medical and industrial applications. Despite its advantages, EIT encounters two primary challenges: the ill-posed nature of its inverse problem and the spatially variable, location-dependent sensitivity distribution. Traditional model-based methods mitigate ill-posedness through regularization but overlook sensitivity variability, while supervised deep learning approaches require extensive training data and lack generalization. 
Recent developments in neural fields have introduced implicit regularization techniques for image reconstruction, but these methods typically neglect the physical principles underlying EIT, thus limiting their effectiveness. 
In this study, we propose PhyNC (Physics-driven Neural Compensation), an unsupervised deep learning framework that incorporates the physical principles of EIT. PhyNC addresses both the ill-posed inverse problem and the sensitivity distribution by dynamically allocating neural representational capacity to regions with lower sensitivity, ensuring accurate and balanced conductivity reconstructions. 
Extensive evaluations on both simulated and experimental data demonstrate that PhyNC outperforms existing methods in terms of detail preservation and artifact resistance, particularly in low-sensitivity regions. Our approach enhances the robustness of EIT reconstructions and provides a flexible framework that can be adapted to other imaging modalities with similar challenges.
\end{abstract}


\section{Introduction}
Non-invasive, portable, and accurate tomographic imaging holds great significance for medical and industrial applications \cite{qi2024bridging}, \cite{lin2022soft}. Though imaging techniques such as computed tomography (CT) can achieve precise imaging \cite{wang2008outlook}, they are constrained by workspace limitations. In contrast, Electrical Impedance Tomography (EIT) enables convenient, non-invasive imaging by estimating the internal conductivity distribution from boundary voltage measurements \cite{adler2021electrical}. Owing to these advantages, EIT has found extensive use in various medical contexts, including bioelectronic medicine \cite{ravagli2020imaging}, \cite{ravagli2021fascicle}, medical imaging \cite{halter2008broadband}, \cite{ murphy2016absolute}, \cite{mahara20153d}, \cite{borsic2009vivo}, \cite{clay2002weighted}, and the monitoring of muscle engagement \cite{zhu2022monitoring}, \cite{farid2022diminished}, \cite{yang2024wearable}. Beyond healthcare, EIT plays an important role in physical human-computer interaction, enabling hand pose estimation \cite{zhang2016advancing}, \cite{kyu2024eitpose}, \cite{zhang2015tomo} and the development of soft robotic skin \cite{tawil2011improved}, \cite{park2022biomimetic}, \cite{chen2024enhancing}, \cite{silvera2014electrical}, \cite{xin2023electrical}, \cite{lee2019internal}, \cite{yang2024body}, \cite{park2021deep}, \cite{park2022neural}. Despite its broad utility, EIT is hindered by two primary {\it challenges} \cite{lin2022soft}, \cite{adler2021electrical}: 
    (i) The ill-posed nature of its underlying inverse problem,
     and (ii) The location-dependent, non-uniform sensitivity distribution \cite{chen2021location} stemming from the properties of the applied current.

Traditional model-based methods address the ill-posed nature of inverse problems by constraining the solution space through regularization techniques (e.g., sparse regularization \cite{jin2012reconstruction} and total variation regularization \cite{borsic2009vivo}). Despite these strategies, their modeling capabilities remain limited, because they do not address the second challenge. Recently, deep learning has emerged as a promising alternative to model-based approaches \cite{zhu2018image}. Leveraging data-driven priors, supervised deep learning (DL) methods have achieved state-of-the-art performance in ill-posed tomographic imaging \cite{wang2020deep}. In the context of EIT, these methods can directly estimate internal conductivity distributions from boundary voltage measurements by training neural networks to learn nonlinear inverse operators \cite{10lecun2015deep}, \cite{11tan2018image}, \cite{12hamilton2018deep}, \cite{13herzberg2021graph}, \cite{14pokkunuru2023improved}, \cite{15chen2020electrical}, \cite{16capps2020reconstruction}. Nonetheless, these methods require large amounts of training data, which is often scarce or expensive to obtain, and they exhibit limited generalization capabilities since they do not directly tackle EIT’s inherent challenges, relying instead on end-to-end mappings.

Neural field approaches represent a novel DL strategy that has shown substantial promise for image reconstruction \cite{ruckert2022neat}, \cite{zang2021intratomo}, \cite{sun2021coil}, \cite{wang2023unsupervised}, \cite{liu2024grapheit}, \cite{reed2021dynamic}, \cite{wu2023unsupervised}, \cite{huang2024high}, \cite{kang2024coordinate}. 
These methods train a multilayer perceptron (MLP) to represent the reconstructed image as a continuous function that maps the spatial coordinates to the corresponding values. A similar approach, the deep image prior (DIP) \cite{ulyanov2018deep}, has also been applied to image reconstruction tasks in EIT \cite{liu2023deepeit}, \cite{xia2024powered} as well as in CT \cite{baguer2020computed} and positron emission tomography (PET) \cite{shan2023deep}, \cite{gong2018pet}. DIP is based on the assumption that a neural network’s architecture can act as a form of regularization, thereby enabling the solution of inverse problems without requiring training data. Similarly, the neural field enforces intrinsic image consistency, providing a strong implicit regularization that constrains the solution space in an under-determined setting \cite{rahaman2019spectral}. Notably, while DIP often suffers from spectral bias \cite{rahaman2019spectral}, \cite{shi2022measuring}, neural field has been shown to mitigate this issue \cite{tancik2020fourier}, thereby improving the capture of high-frequency details. However, despite effectively addressing the ill-posedness of the inverse problem through implicit regularization, existing neural field-based approaches predominantly adhere to computer vision paradigms while neglecting the intrinsic physical properties inherent to the second fundamental challenge of EIT.

In this work, we introduce \textbf{PhyNC} (\textbf{Phy}sics-Driven \textbf{N}eural \textbf{C}ompensation), an unsupervised deep learning framework grounded in EIT’s physical principles. PhyNC addresses both the ill-posed nature of EIT and its non-uniform sensitivity distribution by introducing a neural compensation strategy: it allocates greater representational capacity to coordinates that exhibit lower physical sensitivity, thus balancing the overall sensitivity distribution. By combining the inherent regularization of implicit neural representation with a fidelity constraint derived from the EIT model, PhyNC holds great potential to achieve a more comprehensive and accurate reconstruction than previous approaches. 
To mitigate the model's susceptibility to overfitting, we design a hybrid neural representation that synergistically integrates complementary encoding schemes. This approach is further enhanced by frequency-domain regularization and stochastic sampling strategies, which introduce controlled parameter perturbations during optimization to systematically improve robustness and generalization capability.

To further evaluate our approach, we developed a Hybhash-based reconstruction method as a comparative technique—introduced here for the first time in the context of EIT—to underscore the advantages of PhyNC. We also provide a detailed analysis of the impact of location-dependent sensitivity on reconstruction quality and examine the effects of finite element (FE) mesh discretization by varying the density of FE nodes and elements. Experimental results on both simulated and real data demonstrate that PhyNC not only reproduces shape details with high fidelity but also robustly resists artifacts across regions of varying sensitivity. Moreover, our analysis reveals that the choice of embedding space is crucial to the success of the proposed method, suggesting that similar neural compensation techniques could have broader applications.
Our main contributions are summarized as follows: 
\begin{itemize} 
\item We propose PhyNC, an unsupervised deep learning framework that explicitly integrates the physical principles of EIT to address both the inverse problem’s ill-posedness and the non-uniform sensitivity distribution. To the best of our knowledge, this is the first framework of its kind, and it can be readily adapted to other imaging modalities with analogous properties. 

\item We develop a hybrid representation that harnesses the complementary strengths of multiple neural representation methods. Our comprehensive analysis not only deepens the understanding of embedding-domain neural representations but also provides valuable insights for designing optimal neural representation strategies.

\item We propose a novel training scheme based on random sampling and frequency regularization, which significantly enhances the robustness and accuracy of our encoding method across varying spatial scales.
\end{itemize}

The rest of this paper is organized as follows. Section \ref{sec:2} briefly revisits the mathematical model of EIT and reviews existing neural field works for image reconstruction. Section \ref{sec:method} proposes a EIT reconstruction method with neural compensation strategy. Experimental results are systematically presented in Section \ref{sec:exp}, followed by a comprehensive analysis of outcomes. Finally, Sections \ref{sec:discussion} and \ref{sec:conclusion} offer discussions and conclusions, respectively.

\section{Background} \label{sec:2}
\subsection{Mathematical Model of EIT}
EIT fundamentally comprises dual components of forward and inverse problems. The inverse problem (corresponding to the EIT imaging process) is typically solved indirectly by iteratively solving the forward problem to refine conductivity estimates. The forward problem requires computing boundary voltage measurements for a given conductivity distribution, a process conventionally described by the Complete Electrode Model (CEM) \cite{somersalo1992existence}. For numerical implementation, the CEM is commonly discretized using the Finite Element Method (FEM) \cite{vauhkonen1999three}, where the continuous domain is partitioned into finite elements to approximate the governing partial differential equations. Under the assumption of additive Gaussian noise, the observation model can be expressed as:

\begin{equation}
V = A(\sigma) + \epsilon,
\end{equation}
where \( V \) represents the measured voltages, \(\sigma\) is the conductivity within given domain $\Omega$, \( A(\sigma) \) is the forward solution via FEM, and \( \epsilon \) is the additive Gaussian noise. Consequently, the inverse problem is solved by searching the optimal conductivity \( \hat{\sigma} \) to minimize the discrepancy between the predicted and measured voltages:
\begin{equation}
    \label{m1}
    \hat{\sigma} = \mathop{\mathrm{arg\,min}}\limits_{\sigma} \left\{ \Vert V - A(\sigma) \Vert_2^2 + \alpha R(\sigma) \right\}.
\end{equation}
where \(R(\sigma)\) is the regularization term (typically chosen as total variation \cite{5liu2018image}) introduced to mitigate the ill-posedness of the inverse problem, and \(\alpha\) denotes the regularization weight that balances the data fidelity term and the prior knowledge. 

\begin{figure*}
    \centering
    \includegraphics[width=6.4in]{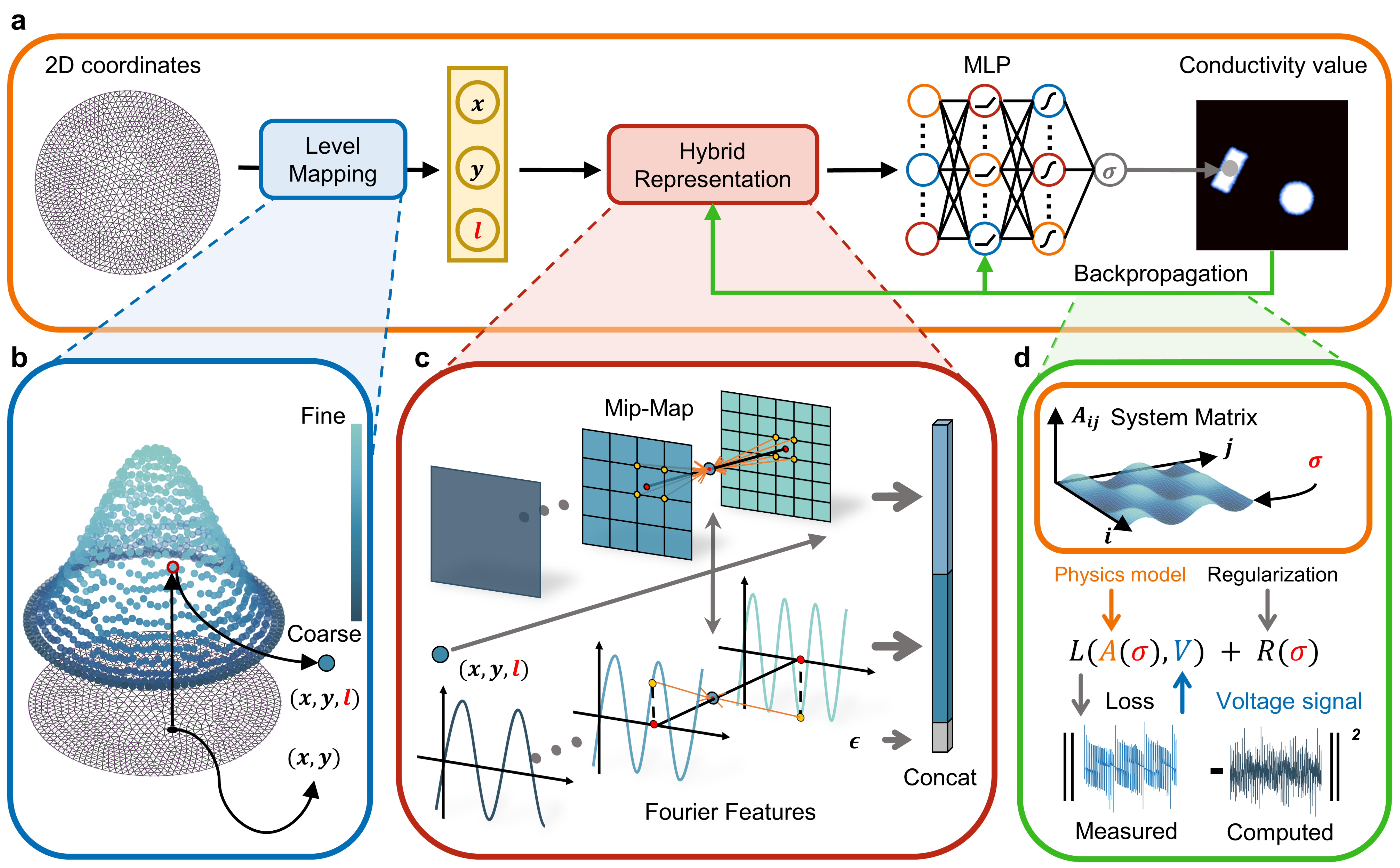}
\caption{\textbf{Overview of PhyNC.} \textbf{a}, The PhyNC model learns a continuous mapping from coordinates to their corresponding conductivity values using a neural field parameterized by an MLP and Hybrid representation. \textbf{b}, Each input coordinate \((x, y)\) is assigned to a specific level \(l\) based on its intrinsic physical properties, forming the triplet \((x, y, l)\). \textbf{c}, Mip-Map embeddings are computed using grids with corresponding resolutions at each level. In parallel, Fourier feature projections are applied at each level with the appropriate frequencies. Addtionally, an global feature is concatenated with the aforementioned two components to enhance stability during reconstruction. \textbf{d}, The physics model projects the conductivity map onto a voltage signal. The loss is calculated between the measured data and the computed voltage to optimize the parameters of the PhyNC model.}\label{fig1}
\end{figure*}

\subsection{Neural Field}
\textit{1) Neural Field for EIT:} Neural Radiance Fields (NeRF) \cite{18mildenhall2021nerf} establishes a continuous implicit representation which learns a mapping from spatial coordinates to target physical quantities. The architecture integrates two key components: a position encoding scheme and a cascade of MLP. Crucially, NeRF avoids direct supervision of MLP outputs. It instead applies task-specific forward operators (e.g., rendering integrals or physical solvers) to map MLP predictions to measurable quantities, then optimizes the model by minimizing simulated-measured discrepancies, enabling unsupervised learning without explicit priors. When adapted to EIT, the NeRF paradigm parameterizes the conductivity field as a continuous implicit function:

\begin{equation}
    \label{eq:nerf_forward}
    \sigma(\mathbf{x}) = f_{\boldsymbol{\Theta}}\left( \gamma(\mathbf{x}) \right)
\end{equation}
where \(\mathbf{x} \in \Omega \subset \mathbb{R}^d\ (d=2,3)\) denotes spatial coordinates within the imaging domain \(\Omega\), \(\gamma(\cdot)\) implements the position encoding operator, and \(\boldsymbol{\Theta}\) represents the MLP’s trainable parameters. In each iteration, the forward operator \(A\left( \cdot \right)\) is applied and the inverse problem is solved via a regularized nonlinear least-squares minimization:

\begin{equation}
    \label{eq:inverse_opt}
    \hat{\boldsymbol{\Theta}} = \underset{\boldsymbol{\Theta}}{\arg\min} \left\{
    \left\| \mathbf{V} - A\left( f_{\boldsymbol{\Theta}} \right) \right\|_{2}^2 + 
    \alpha \mathcal{R}(f_{\boldsymbol{\Theta}})
    \right\}
\end{equation}

\textit{2) Position Encoding Schemes in Neural Field:} 
Positional encoding (PE) serves as a critical preprocessing step to address the spectral bias of coordinate-based MLPs in neural field representations. By mapping low-dimensional spatial coordinates into a high-dimensional feature space through deterministic encoding functions, PE enables MLPs to better approximate high-frequency signals in target scenes. Fourier Feature Projection (FFP) \cite{tancik2020fourier} has emerged as a foundational approach, where input coordinates are transformed using sinusoidal functions of varying frequencies. This frequency-based encoding expands the MLP's receptive field in the spectral domain, effectively mitigating its tendency to prioritize low-frequency components during regression tasks.

Recent advancements introduce learnable encoding strategies to optimize spatial-frequency trade-offs. Hash \cite{26muller2022instant} employs multi-resolution hash tables with trainable feature vectors, enabling efficient memory-computation balance through spatial parameter hashing. Concurrently, Mip-Map \cite{22barron2021mip} extends PE to volumetric rendering by integrating conical frustum approximations and integrated positional encoding, which anti-aliases high-frequency features across scale variations. These innovations address critical limitations of fixed-bandwidth FFP in handling multi-scale details and view-dependent effects in original rendering tasks.

\section{Physics Driven Neural Compensation} \label{sec:method}
In this section, we will first provide a brief introduction to the PhyNC framework. Then, the details will be introduced, including sensitivity based level mapping in Section \ref{tp2}, hybrid representations in Section \ref{tp3}, and frequency regularization in Section \ref{tp4}.
\subsection{Overview}
Fig.\ref{fig1} presents an overview of the proposed PhyNC algorithm, an unsupervised deep learning approach for EIT. The PhyNC pipeline comprises four main components: Level Mapping, Hybrid Representation, an MLP, and the Physics model \textit{A} (i.e., the numerical solution of EIT forward problem). The process begins by taking as input the coordinates of the vertices from a FE mesh, which serves as the foundation for solving the EIT forward problem (Fig.~\ref{fig1}(a)). PhyNC then integrates hybrid representation with the MLP to learn a continuous mapping from coordinates to the corresponding conductivity values. Specifically, each coordinate is assigned to a distinct level based on its intrinsic physical properties relevant to EIT (Fig.~\ref{fig1}(b)). These coordinates are subsequently encoded into a higher-dimensional representation space by combining Mip-Map \cite{22barron2021mip} embeddings and Fourier frequency projections (Fig.~\ref{fig1}(c)). Finally, the resulting hybrid representation is fed into the MLP, producing the conductivity distribution for the target domain.

To address the absence of ground truth for the conductivity distribution, the physics model \textit{A} is employed to transform the conductivity distribution into a predicted voltage signal (Fig.~\ref{fig1}(d)). The loss function \( L \) is calculated between the predicted signals \( A(\sigma) \) and the noisy measured voltage signals \( V \). The PhyNC model then optimizes its parameters through backpropagated gradients, embodying an unsupervised learning paradigm. This approach allows PhyNC to learn directly from the input FE coordinates and corresponding voltage measurements without needing labeled examples, distinguishing it from existing supervised machine-learning methods. Further architectural and implementation details are provided in following subsections.

\subsection{Sensitivity-Based Level Mapping} \label{tp2}
In EIT, the sensitivity matrix quantifies the influence of conductivity perturbations within the imaging domain on boundary voltage measurements \cite{chen2021location}, \cite{alirezaei2009tactile}. This fundamental relationship is expressed through the sensitivity equation:

\begin{equation}
    \frac{\partial A}{\partial \sigma(x,y)} \propto -\|\nabla \phi(x, y)\|^2
    \label{eq:sensitivity}
\end{equation}
where $\phi(x, y)$ denotes the potential distribution in the domain $\Omega$, $\sigma(x,y)$ represents the conductivity distribution.

To isolate the intrinsic sensitivity characteristics of the electrode configuration from conductivity variations, we consider a homogeneous medium with uniform conductivity $\sigma_0$. For a current injection pair at angular positions $\theta_{\text{in}}$ and $\theta_{\text{out}}$, the current density $\mathbf{J}$ and electric field $\mathbf{E}$ satisfy Ohm's law:

\begin{equation}
    \mathbf{J} = \sigma_0 \mathbf{E} = -\sigma_0\nabla\phi
    \label{eq:constitutive}
\end{equation}

Under this homogeneous assumption, the governing equation \cite{liu2023deepeit} reduces to Laplace's equation:
\begin{equation}
    \nabla^2 \phi = 0
    \label{eq:laplace}
\end{equation}

Expressed in polar coordinates $(r,\theta)$ for a circular domain:
\begin{equation}
    \frac{1}{r}\frac{\partial}{\partial r}\left(r\frac{\partial \phi}{\partial r}\right) + \frac{1}{r^2}\frac{\partial^2 \phi}{\partial \theta^2} = 0
    \label{eq:polar_laplace}
\end{equation}

The general solution via separation of variables yields:
\begin{equation}
    \phi(r,\theta) = \sum_{n=1}^\infty r^n \left( B_n \cos n\theta + C_n \sin n\theta \right)
    \label{eq:general_solution}
\end{equation}

Boundary conditions at radius $a$ enforce current injection/extraction through Dirac delta functions:
\begin{equation}
    J_r(a,\theta) = \frac{I}{a} \left[ \delta(\theta - \theta_{\text{in}}) - \delta(\theta - \theta_{\text{out}}) \right]
    \label{eq:bc_current}
\end{equation}

Expanding the delta functions via Fourier series:
\begin{equation}
    \delta(\theta - \theta_0) = \frac{1}{2\pi} + \frac{1}{\pi} \sum_{n=1}^\infty \cos n(\theta - \theta_0)
    \label{eq:delta_expansion}
\end{equation}

Matching coefficients in Equations (\ref{eq:general_solution}) and (\ref{eq:bc_current}) determines the series coefficients:
\begin{subequations}
    \begin{align}
        B_n &= -\frac{I}{\pi \sigma_0 n a^n} \left[ \cos n\theta_{\text{in}} - \cos n\theta_{\text{out}} \right] \\
        C_n &= -\frac{I}{\pi \sigma_0 n a^n} \left[ \sin n\theta_{\text{in}} - \sin n\theta_{\text{out}} \right]
    \end{align}
    \label{eq:fourier_coeffs}
\end{subequations}

Substituting these coefficients into Equation (\ref{eq:general_solution}) and employing logarithmic convergence properties yields the closed-form potential distribution:
\begin{equation}
    \phi(r,\theta) = \frac{I}{2\pi \sigma_0} \ln \left( \frac{1 - 2\frac{r}{a}\cos(\theta - \theta_{\text{out}}) + \frac{r}{a}^2}{1 - 2\frac{r}{a}\cos(\theta - \theta_{\text{in}}) + \frac{r}{a}^2} \right)
    \label{eq:closed_form}
\end{equation}

When adjacent excitation pattern was employed for measurement, wherein sensitivity distribution calculations were performed on each neighboring electrode pair followed by averaging. For normalized radius ($a=1$) and the uniform conductivity, the nodal sensitivity $S_i$ over $K$ electrodes becomes (refer to supplementary note for more details):
\begin{subequations}
\begin{align}
    S_i &= -\frac{1}{K}\sum_{k} |\nabla \phi_{k,i}^2| = -\frac{1}{K}\sum_{k=0}^{K-1} \frac{I^2}{4\pi^2} \notag\\
    &\biggl[ \left(\frac{r_i - \cos(\theta_i - \theta_{k+1})}{D_{k,i}} - \frac{r_i - \cos(\theta_i - \theta_k)}{E_{k,i}} \right)^2 \notag\\
    & + \left(\frac{\sin(\theta_i - \theta_{k+1})}{D_{k,i}} - \frac{\sin(\theta_i - \theta_k)}{E_{k,i}} \right)^2 \biggr] \\
    D_{k,i} &= r_i^2 + 1 - 2r_i \cos(\theta_i - \theta_{k+1}) \\
    E_{k,i} &= r_i^2 + 1 - 2r_i \cos(\theta_i - \theta_k) \\
    \theta_k &= \frac{2k\pi}{K}, \quad k = 0,1,\ldots,K-1
\end{align}
\label{eq:sensitivity_calculation}
\end{subequations}
The PhyNC level mapping integrates sensitivity $S_i$ with geometric sparsity $H_i$, where $H_i$ quantifies the mean surrounding element area. The adaptive grid level combines these metrics through:

\begin{equation}
    \text{level}_i = \mu \cdot \log_{10}(S_i) + \nu \cdot (1 - H_i)
    \label{eq:level_mapping}
\end{equation}

where $\mu$ and $\nu$ control the relative weighting of sensitivity and sparsity terms.
    
Extending this framework to neural fields, we formulate coordinate-dependent sensitivity through the chain rule:

\begin{equation}
    S(\mathbf{x}) = \frac{\partial A}{\partial \mathbf{x}} = \frac{\partial A}{\partial \sigma} \cdot \frac{\partial \sigma}{\partial \mathbf{x}}
    \label{eq:coord_sensitivity}
\end{equation}

where $\mathbf{x}$ denotes nodal coordinates, enabling sensitivity-aware adaptation of the neural field representation.

\begin{figure}[t!]
\centering
\includegraphics[width=6.4in]{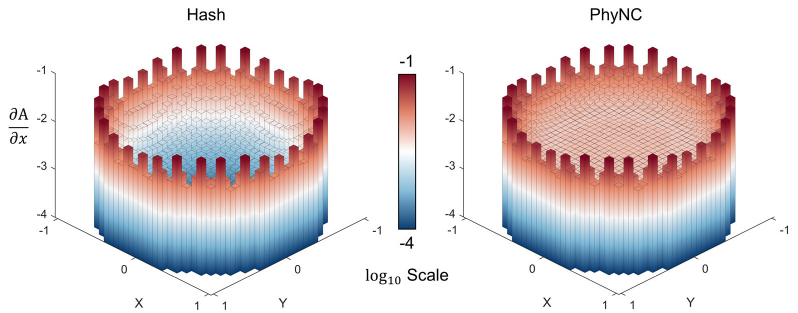}
\caption{\textbf{Toy initial sensitivity visualizations of Hash (left) and our PhyNC (right).} The EIT physics model exhibits location-dependent sensitivity, which diminishes in regions farther from the electrodes. Consequently, Hash—using identical neural representation across all coordinates—naturally inherits this limitation and struggles to capture fine-grained details in these low-sensitivity areas. In contrast, PhyNC assigns higher-sensitivity neural representation to such regions, effectively compensating for the physics model's inherent constraints.}\label{fig:6}
\end{figure}

Building on this definition, we compared PhyNC’s initial sensitivity distribution before iteration with that of Hash \cite{26muller2022instant} to illustrate the impact of neural compensation. In Fig.\ref{fig:6}, we visualize their differences within an uniform conductivity domain. Hash, which encodes all coordinates uniformly, demonstrates markedly lower sensitivity in the central region due to limitations in its underlying physics model. In contrast, PhyNC achieves a more balanced sensitivity distribution through its level mapping process, boosting sensitivity precisely where the physics model traditionally underperforms. However, it is worth noting that as the neural representation of the conductivity \(\sigma(\mathbf{x})\) evolves during iteration, \(\partial \sigma / \partial \mathbf{x}\) (and thus overall sensitivity) may adaptively redistribute. While PhyNC’s initialization mitigates EIT’s inherent sensitivity imbalance, the fidelity of its final reconstructions (especially in low-sensitivity regions) must be empirically validated, as discussed in Section \ref{sec:exp}.

\subsection{Hybrid Representation} \label{tp3}
Inspired by the mipmapping technique in computer graphics \cite{22barron2021mip}, \cite{ 20williams1983pyramidal}, \cite{21hu2023tri}, \cite{23hu2023multiscale}, \cite{24nam2023mip}, we incorporate Mip-Map embedding as one component of our hybrid representation, as illustrated in Fig. \ref{fig1}. This embedding provides a multi-scale representation that captures image details across various resolution levels, effectively balancing the description of coarse structures and fine details.
First, based on the sensitivity distribution of the EIT physics model, we calculate the corresponding level $l$ for each coordinate according to Equation (\ref{eq:level_mapping}). Following this, for each coordinate, position encoding is performed at its corresponding level using both Mip-Map embedding and FFP.

Mip-Map embedding consists of $L$ learnable mipmaps $\mathcal{M}$, where $L$ represents the maximum value of the level. Given a coordinate $(x,y)$ with level $l$, we can get the corresponding feature vectors $m$ from $\mathcal{M}_l$ by trilinear interpolation:
\begin{equation}
    \mathrm{interp}(\mathbf{x},\mathcal{M}_l):(\mathbb{R}^2, \mathbb{R}^{C\times H_l \times W_l}) \rightarrow \mathbb{R}^C ,
\end{equation}
where $C$ is the dimensionality of the feature vectors within the mipmap.
 
 To maintain smooth continuity across non-integer levels $l$, we employ quadrilinear interpolation, which aggregates samples from the two nearest mipmaps (i.e., $\mathcal{M}_{\lfloor l \rfloor}$ and $\mathcal{M}_{\lceil l \rceil}$). The interpolation process can be mathematically expressed as follows:
\begin{equation}
     m = (\lceil l \rceil - l)\mathrm{interp}(\mathbf{x},\mathcal{M}_{\lfloor l \rfloor}) + (l - \lfloor l \rfloor)\mathrm{interp}(\mathbf{x},\mathcal{M}_{\lceil l \rceil}), 
 \end{equation}
where $\lfloor \cdot \rfloor$ and $\lceil \cdot \rceil$ denote the floor and ceiling functions, respectively. This mechanism ensures seamless blending of sampled values between neighboring mipmaps, thereby preserving the continuity of the representation across different levels. The resolution of the mipmaps is determined using the approach from Hash \cite{26muller2022instant}: 
\begin{equation}
    b :=\mathrm{exp}(\frac{\mathrm{ln} N_{max}-\mathrm{ln}N_{min}}{L-1}),
\end{equation}
where $N_{max}$ and $N_{min}$ are the finest and coarsest resolution of mipmaps.

In parallel, for FFP, the coordinates are encoded with frequencies corresponding to their respective levels $l$. We adopt the positional encoding from Wang et al. \cite{wang2023unsupervised} as follows:
\begin{equation}
    f = \mathcal{F}(\mathbf{x}) = [\sin{2\pi B_l \mathbf{x}},~ \cos{2\pi B_l\mathbf{x}}],
    \label{eq:fourier_feat}
\end{equation}
where $B_l$ represents frequencies randomly sampled from the Gaussian distribution $\mathcal{N}(0, s_l^2)$. To accommodate varying frequency ranges across levels, each level $l$ is assigned a distinct variance $s_l$, which governs the bandwidth of frequencies that the neural network can represent 
\begin{equation}
    s_l = s_0 \times \eta^l, \label{freq}
\end{equation}
where $s_0$ is the base variance and $\eta$ is the scaling factor. Finally, the different types of positional encoding are concatenated to form the final feature vector $p$ 
\begin{equation}
    p = \mathrm{concat}(f,m,\xi),
\end{equation}
where $\xi$ denotes the global feature, which serves to enhance background consistency and stabilize the reconstruction process. 

\subsection{Frequency Regularization for Neural Field Optimization} \label{tp4}
The PhyNC framework addresses the inherently ill-posed inverse problem of reconstructing two-dimensional conductivity distributions from severely underdetermined one-dimensional boundary voltage measurements. Although the framework eliminates the need for prior knowledge or pre-trained models, a critical challenge remains: neural networks trained via gradient descent exhibit a pronounced spectral bias, tending to overfit high-frequency components. This behavior, as revealed by spectral bias analyses in Fourier feature mappings \cite{tancik2020fourier}, leads to premature convergence toward high-frequency artifacts that obscure the low-frequency structural details vital for accurate EIT reconstruction.

To alleviate this spectral imbalance, we introduce a frequency regularization strategy inspired by recent advances in neural radiance field optimization \cite{yang2023freenerf}. In contrast to brute-force spectral masking techniques, our method employs progressive frequency augmentation through parametric control of the FFP encoding. Specifically, the base bandwidth parameter, $s_{0}$ (defined in Equation (\ref{freq}) and discussed in \cite{wang2023unsupervised}), is modulated using a tunable sigmoidal transition function given by

\begin{equation} s_0(t) = s_{\text{min}} + \frac{s_{\text{max}} - s_{\text{min}}}{1 + e^{-k(t - s_{\text{th}})}}, \label{eq:param_sigmoid} \end{equation}

where \(t\) is the current iteration step, $s_{\text{min}}$ and $s_{\text{max}}$ denote the lower and upper saturation bounds, respectively. The parameter $s_{\text{th}}$ specifies the critical input value at which the function reaches the median bandwidth, i.e., $\frac{s_{\text{min}}+s_{\text{max}}}{2}$, and the steepness factor $k$ governs the rate of transition; larger values of $k$ induce a sharper transition.

At fixed training intervals $T$, the base bandwidth $s_0$ is recalculated using Equation (\ref{eq:param_sigmoid}). This updated bandwidth parameter is then leveraged to resample $B_l$ through Equation (\ref{eq:fourier_feat}), followed by the computation of new Fourier features to advance the training process. Such spectral regularization, combined with stochastic resampling, introduces controlled perturbations into the coordinate-based neural representations, thereby enhancing the overall model stability and generalization capability.

\begin{figure*}[!t]
\centering
\includegraphics[width=7.12in]{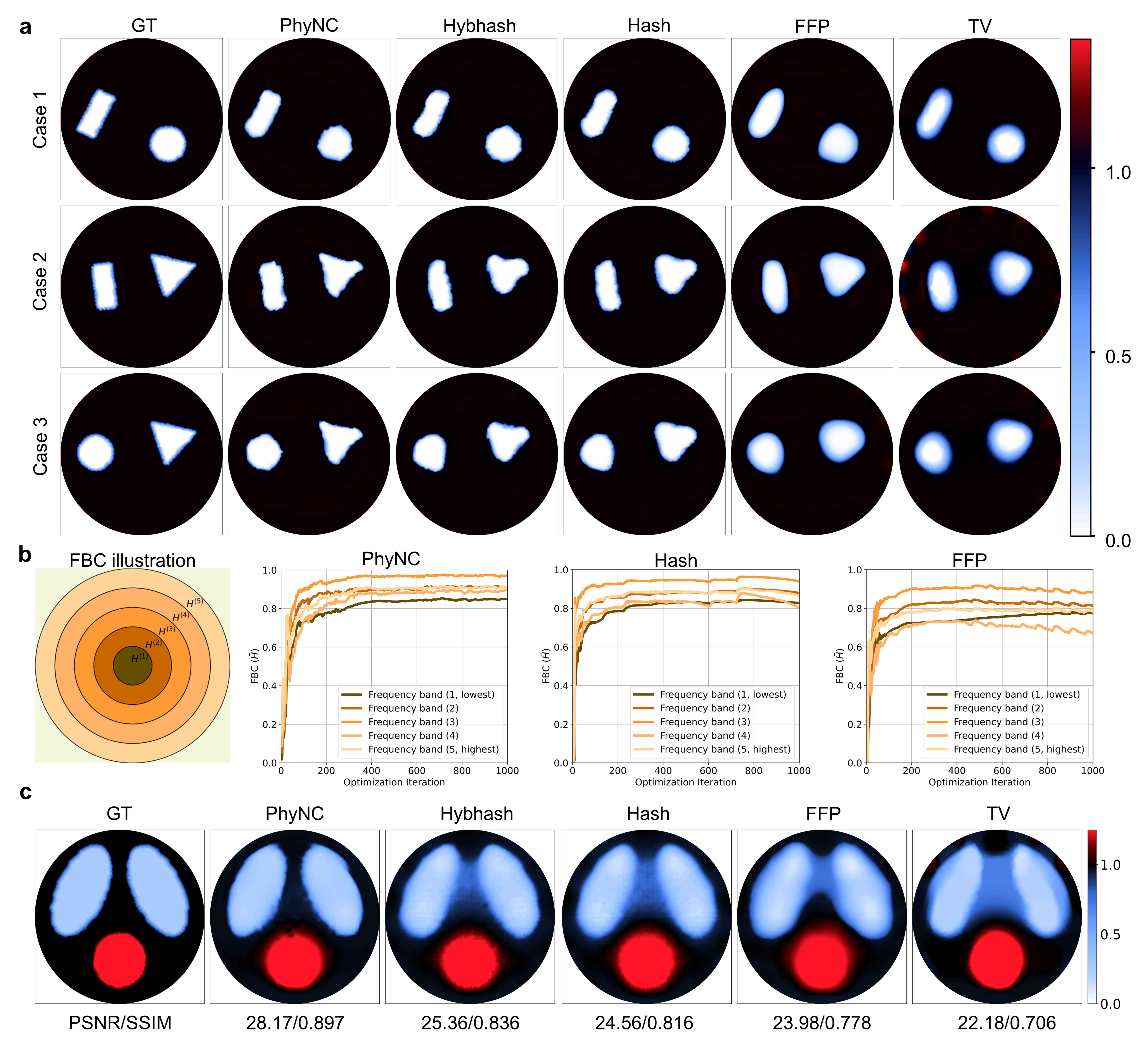}
\caption{\textbf{Results on simulated data.} \textbf{a}, Conductivity maps reconstructed by PhyNC and the baseline methods. Our method effectively recovers local details while maintaining background consistency. \textbf{b}, Spectral measurements comparing the proposed approach with our baselines on case 1 are presented, including illustrations and metrics for Frequency Band Correspondence. \textbf{c}, Reconstruction results on low-contrast simulated data demonstrate that our method effectively eliminates the staircase artifacts in the central area. The color bar is displayed on the right. }
\label{fig2}
\end{figure*}

\begin{table*}[!ht]
\begin{center}
\caption{Quantitative Comparison of Our Method versus baselines with Simulated Data Featuring Various Geometric Shape Combinations}
\label{tab:table1}
\resizebox{0.75\linewidth}{!}{
\begin{tabular}{l | ccc | ccc | ccc} 
\toprule
\multirow{1}{*}{Method} & \multicolumn{3}{c|}{PSNR $\uparrow$}& \multicolumn{3}{c|}{SSIM $\uparrow$}&\multicolumn{3}{c}{LPIPS $\downarrow$}\\ 

& Case 1& Case 2&Case3&  Case 1& Case 2&Case 3& Case 1& Case 2&Case 3\\ 
\midrule
 TV& 22.39& 19.79& 20.63& 0.887& 0.779& 0.828& 0.046& 0.122&0.078\\
 FFP& 24.10& 22.34& 21.11& 0.903& 0.878& 0.857& 0.021& 0.029&0.037\\
 Hash& 26.81& 25.05& 24.62& 0.936& 0.922& 0.922& 0.012& 0.024&0.020\\
 Hybhash& 28.05& 23.89& 24.23& 0.945& 0.919& 0.922& 0.011& 0.025&0.020\\
 PhyNC& \textbf{28.30}& \textbf{27.73}& \textbf{29.28}& \textbf{0.952}& \textbf{0.944}& \textbf{0.958}& \textbf{0.011}& \textbf{0.019}&\textbf{0.005}\\
\bottomrule
\end{tabular}
}    
\end{center}
\end{table*}

\section{Experiments and Results} \label{sec:exp}
\subsection{Experimental Setup}
\subsubsection{Data} We assessed our method through simulations and experimental tests. In the simulations, we conducted three cases within a normalized circular domain (radius 1 cm) using 16 evenly spaced electrodes. We injected 1 mA currents between selected pairs of electrodes, totaling 54 unique injections, and measured voltages between adjacent pairs. The EIT data was generated using a dense finite element mesh (5,833 nodes, 11,424 elements) with added Gaussian noise (SNR = 60 dB). Experimentally, various plastic objects were submerged in a saltwater tank to create different conductivity distributions, mirroring the simulation setup. Data was collected using the KIT-4 measurement system \cite{27kourunen2008suitability}. To further verify the effectiveness of the proposed method, we designed low-contrast cases where the inclusions had conductivity values closer to that of the background. To prevent inverse crime and accelerate the iteration process, we performed the inversion on a coarser mesh (1,145 nodes, 2,176 elements) and tested on a finer mesh (5,833 nodes, 11,424 elements), demonstrating the method's generalization capacity. 

\subsubsection{Baselines} We compare our method against four established baselines specifically designed for this task: \textbf{TV}: A traditional physics model based on Total Variation (TV) regularization \cite{5liu2018image}; \textbf{FFP}: An approach that first employs an implicit neural field for EIT and uses a Fourier frequency function to encode the coordinates \cite{wang2023unsupervised}; \textbf{Hash}: A method that leverages grid-based spatial partitioning and utilizes hash mapping to manage memory consumption \cite{26muller2022instant}; \textbf{Hybhash}: A hybrid approach that developed by us alongside PhyNC, integrating FFP and Hash embedding to encode the input coordinates. For consistency across all methods, we use the official implementations provided by the authors. 
However, we adjust the model capacity (number of parameters and grid/hash table size) and task-specific parameters to ensure a fair comparison. In particular, for Hash and Hybhash, we set the maximum hash table size to \(T = 2^{17}\) and the grid level to \(L = 32\). Additionally, the FFP dimension used in Hybhash is set to 16. 

\subsubsection{Implementation Details}
We employed the AdamW optimizer for 1,000 iterations, initializing the learning rate at \(5 \times 10^{-3}\) for the MLP and \(5 \times 10^{-2}\) for the Mip-Map features. The learning rate was subsequently adjusted using a cosine annealing schedule. The MLP architecture comprised four layers, each containing 128 units. Rectified Linear Unit (ReLU) activation functions were used between the fully connected layers, while the final layer employed a Sigmoid activation function followed by a scale factor multiplication. The mipmap was configured with 16 levels, ranging from the finest resolution of 64 to the coarsest resolution of 4. The feature dimensions for Mip-Map and FFP is 32, and the global feature is 16. All computational tasks were executed on a system equipped with an NVIDIA RTX 4070 GPU and an Intel i7-12700K CPU.

\subsection{Detail recovery and Anti-artifacts by PhyNC}
We first evaluated the performance of PhyNC using simulated data. Specifically, we tested the method on reconstructions of three basic geometric shapes: a circle, a rectangle, and a triangle. The reconstruction results are shown in Fig. \ref{fig2}(a). PhyNC effectively reconstructs each object's shape, size, and position, with notable accuracy in capturing the right angles of rectangles and the acute angles of triangles. Moreover, the proposed method shows superior image quality compared to reference methods, as confirmed by quantitative evaluation metrics with results summarized in Table \ref{tab:table1}.

Fig. \ref{fig2}(b) provides further insights into the performance of PhyNC, where we compare the reconstructions of FFP, Hash and PhyNC across various frequency bands. At lower frequencies, all two embedding-based methods—Hash, and PhyNC—show improved convergence and accuracy, as evidenced by their high correspondence with the ground truth. However, in terms of the Frequency Band Consistency (FBC) values \cite{shi2022measuring}, our method consistently outperforms the others, particularly in the higher frequency bands. This superior performance is also reflected in the reconstructed images shown in the second column from the left of Fig. \ref{fig2}(a), where PhyNC demonstrates better preservation of fine details and more accurate inclusion recovery. To further evaluate the robustness of the method, we conducted a low-contrast experiment using a simulated heart-and-lungs phantom (see supplementary Fig. 2 for additional low-contrast case). PhyNC provides an almost perfect reconstruction of the target shape, achieving the highest corresponding metrics, as shown in Fig. \ref{fig2}(c). In contrast, the reference methods introduce notable artifacts, such as blurred edges and staircase-like effects, which hinder the accurate delineation of boundaries between the target and surrounding tissues. These artifacts are especially problematic in low-contrast regions, where precise reconstruction is essential for identifying subtle anatomical features.

\subsection{Sensitivity compensation and Reconstruction robustness by PhyNC}
\begin{figure*}[!t]
\centering
\includegraphics[width=6.5in]{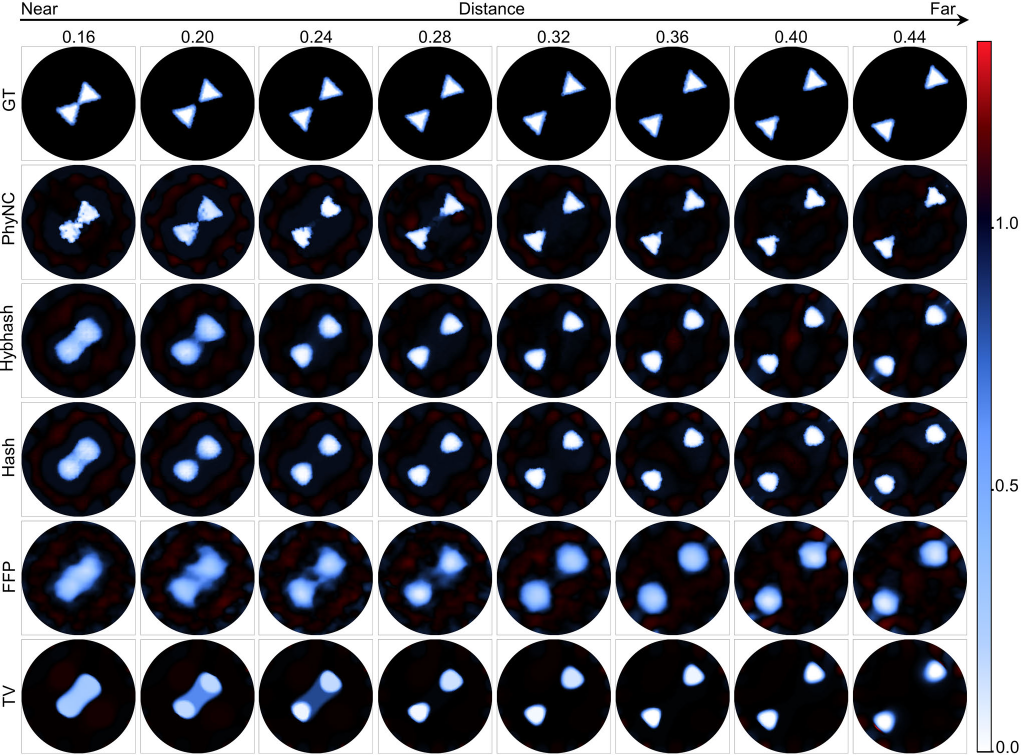}
\caption{\textbf{Location-dependent sensitivity effects on reconstruction performance.} Reconstruction results for two small triangles at varying locations from the center.}\label{fig:3a}
\end{figure*}

\begin{figure*}[!t]
\centering
\includegraphics[width=5.72in]{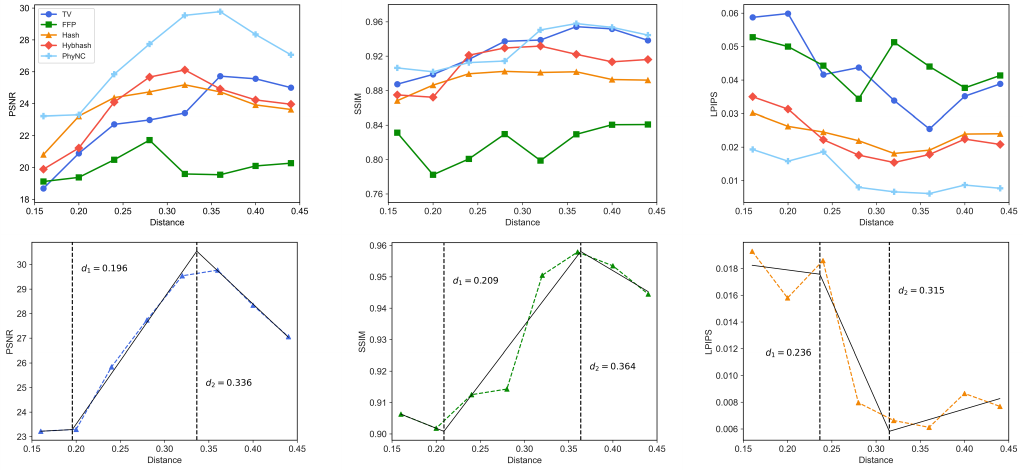}
\caption{\textbf{Quantitative evaluation.} The top row presents the reconstruction quality across all methods at each location using PSNR, SSIM, and LPIPS metrics. The bottom row displays three-segment piecewise linear fits for PhyNC (solid black lines), which delineate two critical distance thresholds (vertical dashed black lines).}\label{fig:3b}
\end{figure*}

Although PhyNC has demonstrated initial success in reducing artifacts and enhancing details in simulated data, the practical application of EIT remains constrained by the highly inhomogeneous, location-dependent sensitivity distribution. This issue is particularly pronounced when the target object is small and located in deep regions of the imaging domain, where sensitivity to conductivity changes is minimal due to low potential and current density. Therefore, we validate PhyNC's ability to enhance sensitivity and examine its specific impact on reconstruction quality before applying the method to experimental data.

To explore the impact of the location-dependent sensitivity, we manipulated the normalized distance (dimensionless and without physical units) between the target objects and the center of the measurement domain. Specifically, we used two small triangular objects with a side length of 0.4, acquiring images at increasing sensitivity levels as the distance between the objects and the center was increased (Fig. \ref{fig:3a}). At very short distances (e.g., 0.16, as shown in the first column of Fig. \ref{fig:3a}), the sensitivity was insufficient for clear and accurate reconstruction. However, our proposed PhyNC, while still struggling to achieve perfect reconstruction in the low-sensitivity region, was able to roughly distinguish between the two objects and capture their contours. When the distance from the center to the objects was increased to 0.20 (column 2, Fig. \ref{fig:3a}), PhyNC performed significantly better, successfully distinguishing the two objects and reproducing their shape details. This was a marked improvement over the other methods, particularly in the absence of staircase artifacts. As the distance increased further to approximately 0.4 (column 4-7, Fig. \ref{fig:3a}), the shape details and conductivity values of the triangles were consistently well restored by PhyNC, demonstrating its robust performance. Interestingly, when the distance became too large (e.g., 0.44, as seen in the last column of Fig. \ref{fig:3a}), the reconstruction performance began to deteriorate. This decline is likely due to the drastic change in sensitivity as the objects are placed too close to the electrodes, which makes the solution space more complex and harder for the model to learn accurately.

\begin{figure*}[!t]
\centering
\includegraphics[width=7.12in]{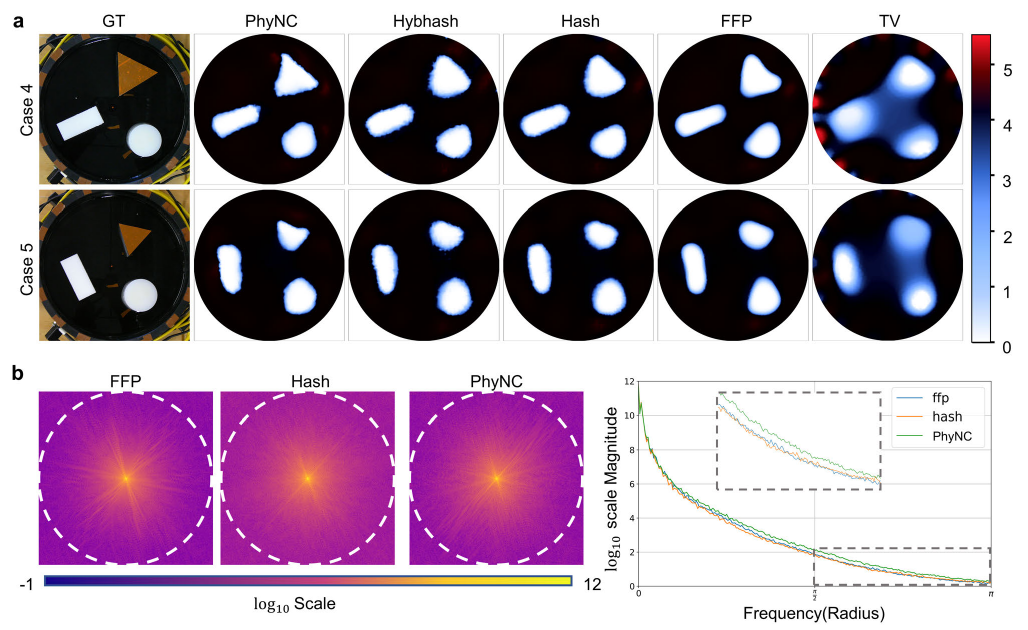}
\caption{\textbf{Results on experimental data.} \textbf{a,} Conductivity maps reconstructed by PhyNC and the baseline methods. \textbf{b}, Spatial-frequency representation of images in Case 4 and their radially averaged profiles.}\label{fig:4a}
\end{figure*}

\begin{figure}[!ht]
\centering
\includegraphics[width=3.5in]{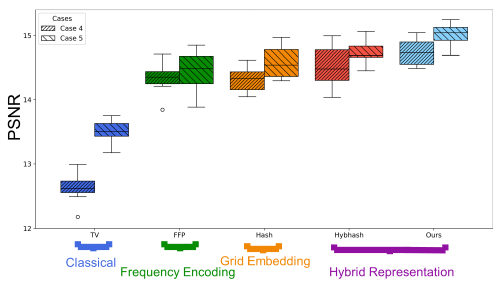}
\caption{The calculated PSNR and SSIM on Case 4 and Case 5 under five different random seed tests.}\label{fig:4b}
\end{figure}

To quantify the performance of each method, we utilized three standard metrics: PSNR, SSIM \cite{wang2004image}, and LPIPS \cite{zhang2018unreasonable}. As expected, performance improved significantly with increasing distance, as visually observed (the top row of Fig. \ref{fig:3b}). However, when the distance became too large, especially when the objects were placed too close to the electrodes, the performance metrics began to degrade. PSNR and SSIM dropped, while LPIPS increased, which is consistent with the visual observations in Fig. \ref{fig:3a}. Using three-segment piecewise linear fits to analyze the PSNR, SSIM, and LPIPS data, we identified a cutoff distance of approximately 0.213 at which PhyNC provides robust structural recovery for the two triangular objects (the bottom row of Fig. \ref{fig:3b}). Beyond this cutoff, the performance of PhyNC remained strong up to a distance of about 0.338. Beyond this threshold, the reconstruction performance began to degrade, likely due to the overly complex sensitivity distribution. PhyNC demonstrates superior performance across all distances, particularly in the low-sensitivity region, where it significantly outperforms other methods. This indicates that PhyNC is highly effective in compensating for sensitivity. 

\subsection{PhyNC for experimental data}
After validating PhyNC using simulated data and assessing its sensitivity compensation performance, we applied the method to experimental water tank data (See more results of the experimental data in Supplementary Figs. 3\&4). The results for Case 4 are presented in the first row of Fig. \ref{fig:4a}. The conventional reconstruction method fails to accurately reconstruct multiple objects, exhibiting severe electrode artifacts and staircase-like effects. In contrast, all neural field-based methods successfully distinguish the inclusions, with PhyNC providing the most detailed reconstruction. It effectively restores the fine details of triangles and rectangles in low-sensitivity regions, where other methods struggle. In Case 5, a smaller plastic triangle was placed, and its position was altered (second row of Fig. \ref{fig:4a}). Once again, PhyNC produced the best imaging quality, yielding clean backgrounds and the most accurate reconstruction of the triangle’s details. It outperformed all other methods in this comparison, demonstrating its superior performance in challenging scenarios.

To further assess the sensitivity improvements, we analyzed the spatial-frequency characteristics of the reconstructions from Case 4 using Fourier transforms. The introduction of embedding-based positional encoding, via an embedding space, resulted in larger magnitudes in the high spatial-frequency range compared to FFP (i.e., away from the origin in the spatial-frequency representation; see left panels of Fig. \ref{fig:4a}(b)). This improvement is also evident from the radially averaged line power spectral density profiles (right panels of Fig. \ref{fig:4a}(b)), which reveal a marked recovery of high spatial-frequency information. When comparing PhyNC to Hash, similar performance was observed at low and medium spatial frequencies. However, PhyNC demonstrates a significant advantage in the high spatial-frequency range (inset in the dashed box, Fig. \ref{fig:4a}(b)). This result suggests that the sensitivity compensation mechanism in PhyNC contributes to finer details restoration, which aligns with visual observations of the reconstructed images.

\begin{figure*}[!t]
\centering
\includegraphics[width=5.72in]{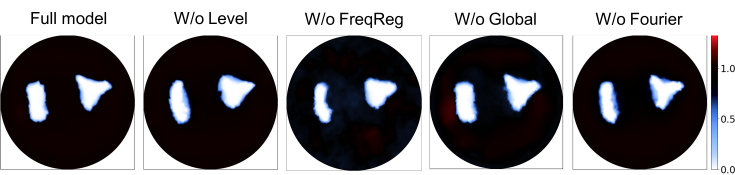}
\caption{\textbf{Ablation study.} We compare our method against its variants with the Fourier features and/or the proposed neural compensation strategy disabled and Frequency regularization strategy.}\label{fig:5a}
\end{figure*}

We then computed the PSNR and SSIM for both Case 4 and Case 5, using five runs with different random seed values to account for the stochastic nature of the training process (Fig. \ref{fig:4b}). The results demonstrate that the neural field-based approach outperforms traditional physical modeling methods. Furthermore, although the embedding-based method (Hash) benefits from the introduction of an embedding space, it exhibits high variance, as indicated by the larger spread in the results (Fig. \ref{fig:4b}). In contrast, the hybrid representation method combines the advantages of both approaches, offering superior performance and greater stability, as reflected by its smaller variance (Hybrid representation, Fig. \ref{fig:4b}). Overall, PhyNC achieves the most robust and optimal results by leveraging hybrid representation and sensitivity compensation based on physical principles.

\subsection{Ablation study and Representation analysis}

We have demonstrated the superior performance of PhyNC and now seek to identify the key components contributing to its effectiveness. To this end, we first conduct an ablation study on Case 2 (Fig.~\ref{fig:5a}) to assess the impact of removing specific components of the model, namely Fourier features, neural compensation, frequency regularization and global features. The results show that removing the global feature causes the background to display abnormally high values at the object’s edges (indicated by the red area in the figure), accompanied by a slightly blurred outline. When the neural compensation is removed, the reconstruction performance significantly deteriorates, as the model begins to encode identical positions for all nodes, losing fine shape details, particularly in low-sensitivity regions. The removal of Fourier features leads to a marked degradation in reconstruction quality, highlighting the critical role of hybrid representation in improving model performance.

\begin{figure}[!t]
\centering
\includegraphics[width=5.6in]{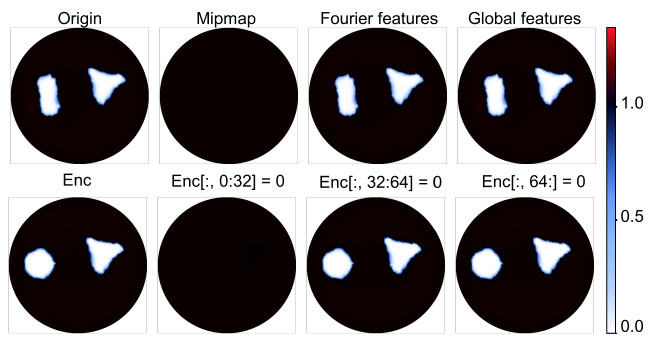}
\caption{The effect of setting each type of representation (Mip-Map embeddings, Fourier features, and global features) to 0, is evaluated sequentially from left to right. This was accomplished by adding a single line of code.}\label{fig:5b}
\end{figure}
After confirming the importance of each model component, we next investigate the influence of different types of positional encoding on image reconstruction in Cases 2 and 3. Specifically, we zero out the outputs of Mip-Map embedding, Fourier features, or global features and show the resulting images in Fig.~\ref{fig:5b}. We observe that all shape details—such as size, position, and outline—are captured in the Mip-Map, and when other types of positional encoding are set to zero, the result remains largely unchanged. In contrast, setting the Mip-Map to zero results in an image consisting almost entirely of a background with no discernible object. These findings demonstrate that the Mip-Map embedding is crucial for capturing high-frequency details in the reconstructed image.
\begin{figure*}[!t]
\centering
\includegraphics[width=6.4in]{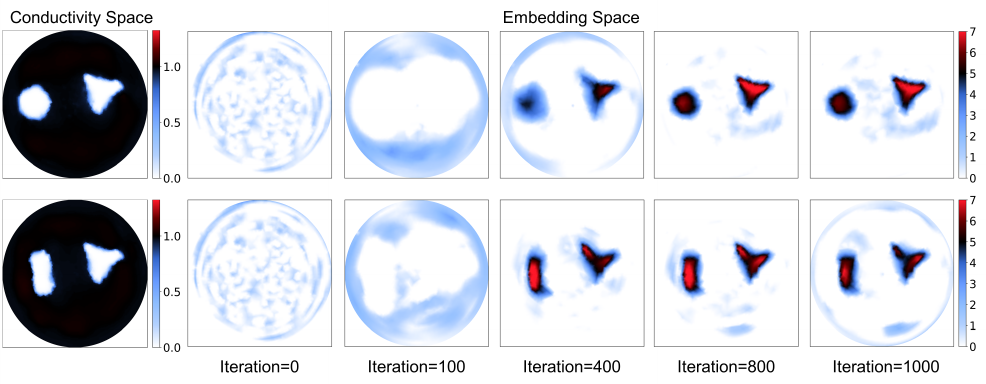}
\caption{\textbf{Principal Component Analysis.} PCA plot of the Mip-Map embedding space, capturing both fine-grained features and significant structural variations for a nuanced representation. Notably, the color bar in the embedding space visualization does not have a direct physical interpretation, whereas, in the conductivity space, it corresponds to actual conductivity values.}\label{fig:5c}
\end{figure*}
Having established the critical role of the Mip-Map embeddings, we next investigate how the embedding space contributes to reconstruction. We perform Principal Component Analysis (PCA) on the Mip-Map embeddings at different stages of training, as shown in Fig.~\ref{fig:5c}. Initially, the embedding space is uniformly initialized, and the PCA output appears as noise. However, as training progresses, the rough outline of the target object gradually emerges. At the end of the training, the embedding space visualization closely aligns with the final conductivity distribution. In contrast, Fourier features rely solely on an MLP to learn the complex conductivity distribution. The learnable Mip-Map embeddings, which can approximate the conductivity distribution throughout training, significantly enhance model performance. This underscores why the Mip-Map embeddings are pivotal to improving the model's reconstruction quality.

\subsection{Trade-Off Between FEM and Neural Field}
\begin{figure*}[!t]
\centering
\includegraphics[width=6.4in]{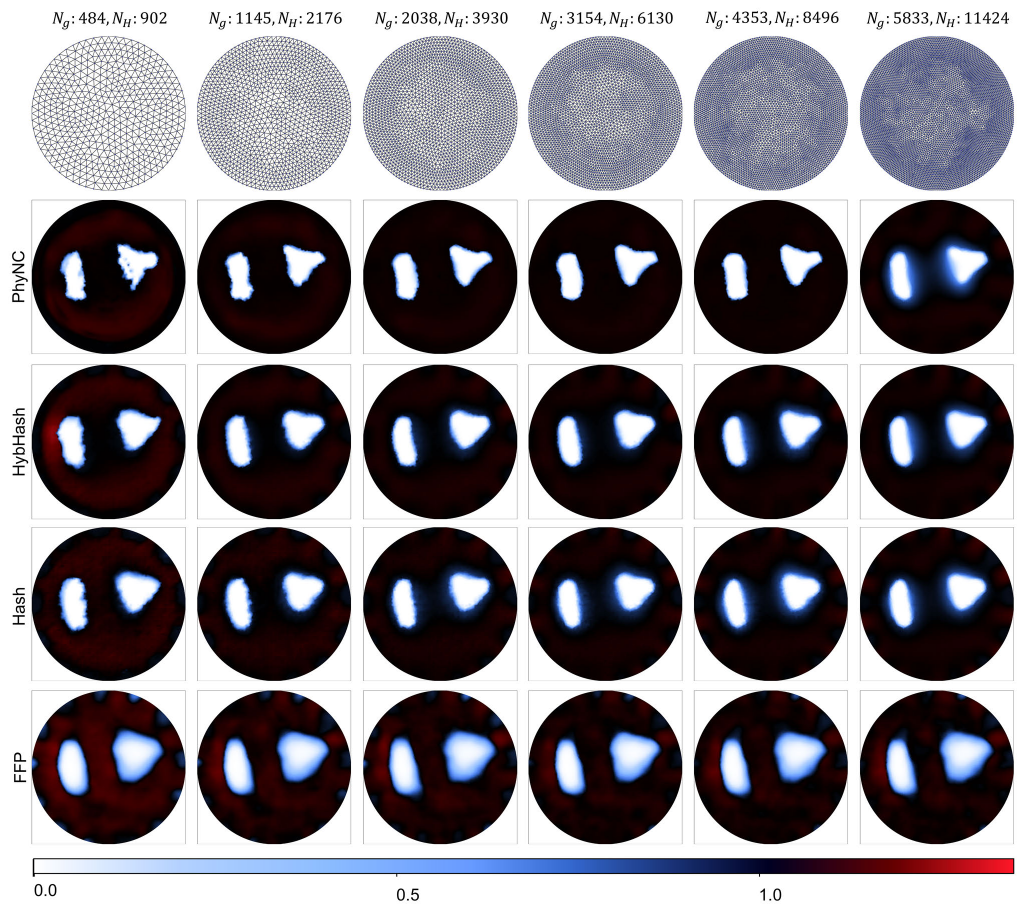}
\caption{\textbf{Results on different meshes.} A comparative analysis of various methods based on training outcomes derived from meshes with differing node densities.}\label{fig:5}
\end{figure*}

Traditional FEMs have long been employed to estimate conductivity values by assigning them to each element within a finely discretized mesh. However, this approach encounters a fundamental challenge: the number of unknown conductivity values typically far exceeds the available measurements, resulting in an excessively large and ill-posed solution space. To mitigate this issue, researchers commonly employ regularization techniques and integrate prior knowledge regarding material properties and anatomical structures. These strategies constrain the solution space, guiding the reconstruction toward conductivity distributions that are both physically meaningful and clinically plausible.

In contrast, neural field methods directly predict conductivity values at each spatial location. In fully supervised settings—where the problem is not inherently underdetermined—augmenting the number of sampling points generally leads to improved performance. However, in EIT, where FEM and neural field methods are integrated, a unique trade-off emerges. Increasing the FEM mesh density (i.e., the discretization level) allows for finer spatial detail capture but simultaneously intensifies the ill-posedness of the inverse problem. This exacerbation can induce issues such as underfitting, overfitting, or training instability in the neural field component.

To systematically investigate the interplay between FEM mesh resolution and model performance, we conducted experiments using six FE meshes with empirically chosen discretization levels. In our study, the number of FE nodes (\(N_g\)) ranged from 484 to 5,833, and the number of elements (\(N_H\)) ranged from 902 to 11,424. Our experimental results (Fig. \ref{fig:5}) reveal several critical insights:
\begin{itemize}
\item \textbf{Underfitting at Low Resolution:} With very few FE nodes, all methods exhibit underfitting due to insufficient spatial detail.
\item \textbf{Improvement with Moderate Resolution:} As the number of nodes increases, reconstruction quality generally improves across all methods.
\item \textbf{Critical Tipping Point:} Beyond a certain threshold—illustrated by the finest mesh (\(N_g=5,833\))—all methods suffer sharp performance degradation, as the exacerbated ill-posedness overwhelms the models’ reconstruction capabilities.
\end{itemize}
These observations underscore the importance of selecting an optimal FEM mesh resolution. Notably, with a moderately fine mesh (\(N_g=3,154\)), most conventional methods begin to decline in performance. In contrast, our proposed PhyNC method continues to improve, reaching peak performance at this discretization level. To quantitatively assess each method's resilience to increasing ill-posedness, we performed two-stage linear fittings on standard image quality metrics—PSNR, SSIM, and LPIPS—to identify a "limiting number of nodes" for each approach. (It is important to note that these limiting values are derived solely from the fitting process and were not directly tested with meshes corresponding to those exact node counts.)
\begin{figure*}[!t]
\centering
\includegraphics[width=6.25in]{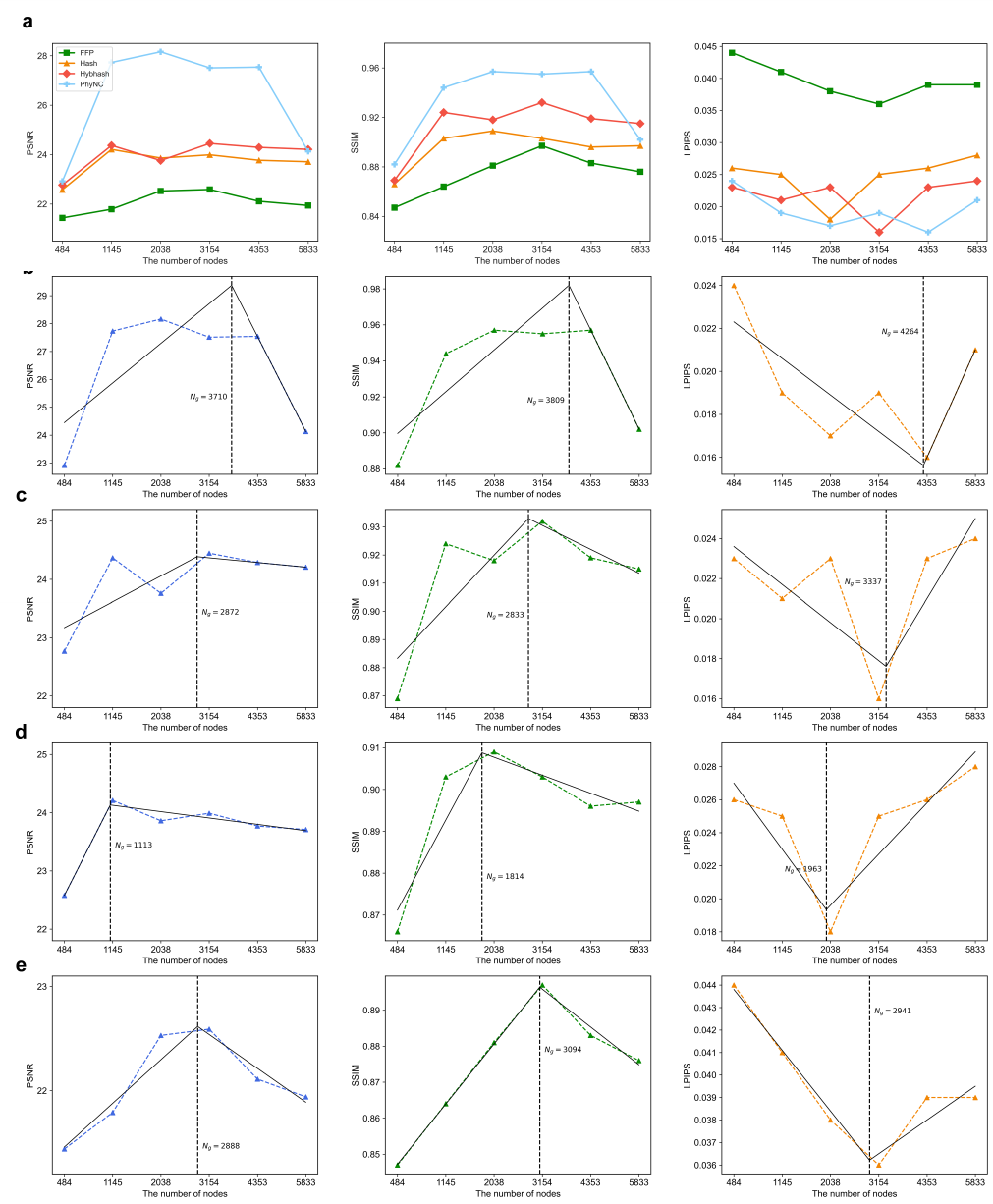}
\caption{\textbf{Quantitative analysis.} \textbf{a,} Quantitative evaluation of reconstruction quality based on PSNR, SSIM, and LPIPS for all methods across varying numbers of nodes.
 \textbf{b-e,} Two-stage linear fitting of PSNR, SSIM, and LPIPS metrics for the PhyNC, Hybhash, Hash and FFP across different node counts.}\label{fig6}
\end{figure*}

As depicted in Fig. \ref{fig6}(a), PhyNC consistently outperforms the other methods under nearly all conditions. Specifically, PhyNC achieves an average limiting node count of 3,928, significantly higher than that of the Hash-based method (1,630 nodes), the FFP method (2,974 nodes), and the hybrid positional encoding variant (Hybhash, 3,014 nodes). This performance hierarchy aligns with theoretical expectations: while hash-based parameterizations excel at reconstructing fine details, they often suffer from stability and generalization issues; FFP methods provide enhanced stability and broader generalization, albeit sometimes at the expense of detail resolution; and the Hybhash method attempts to strike a balance between these extremes. PhyNC, however, offers a distinct advantage by incorporating a neural architecture explicitly designed to account for the physics of EIT. This physics-informed approach embeds an intrinsic prior that effectively guides the reconstruction away from non-physical solutions. Consequently, PhyNC maintains robust stability and achieves high-fidelity reconstructions even under the highly underdetermined conditions imposed by denser FEM meshes.

In summary, our comprehensive study reveals a delicate trade-off in EIT reconstruction. While increasing the FEM discretization level can initially enhance performance by capturing finer spatial details, surpassing an optimal threshold exacerbates the ill-posedness of the inverse problem, ultimately diminishing reconstruction quality. Unlike pure neural field methods—where more sampling points generally lead to improved performance in fully supervised contexts—the hybrid EIT framework necessitates a careful balance between FEM mesh density and neural field training. PhyNC's innovative integration of physical constraints with hybrid representation strategies significantly extends the practical feasibility of EIT reconstructions, outperforming existing methods across multiple metrics. 
\section{Discussion} \label{sec:discussion}
PhyNC introduces a novel DL framework that integrates a neural field with a physical forward model to iteratively extract conductivity information. By compensating for the uneven sensitivity of physical models in neural fields, PhyNC enables highly accurate reconstruction from FE meshes. As an unsupervised learning approach, PhyNC distinguishes itself from traditional supervised methods by eliminating the need for an external training dataset. This characteristic is particularly advantageous, as it allows the model to reconstruct unknows (i.e., conductivity in EIT) on relatively coarse meshes while still generalizing effectively to fine meshes, facilitating high-resolution conductivity distributions. 
A significant advantage of the PhyNC framework lies in its capability to handle rough meshes while maintaining the quality of the final reconstruction. This stands in sharp contrast to other existing methods, which generally rely on using fine meshes to produce high-resolution outcomes.

PhyNC differs fundamentally from 3D reconstruction methods in the field of computer vision that use input points to predict color and density values to render an entire 3D scene. While these methods are elegant and effective in many contexts, they are not ideally suited for applications such as EIT or other tomographic imaging techniques where sensitivity distributions are non-uniform. PhyNC, by contrast, leverages the intrinsic properties of the EIT, performing targeted optimization within the neural field to enhance reconstruction accuracy. Additionally, PhyNC employs a hybrid representation that combines the strengths of both Fourier features and Mip-Map embeddings. This hybrid representation ensures both stability and generalization, ultimately achieving superior performance in terms of reconstruction fidelity.

Finally, it is crucial to highlight that a key factor in PhyNC’s effectiveness is its learnable embedding space. Our analysis reveals that during reconstruction, the embedding space naturally adapts to the numerical distribution of the target space. This adaptive behavior implicitly introduces a structural prior, leading to a significant improvement in model performance.
These insights enhance our understanding of neural representation in neural fields and suggest that this approach may have broader applications in other domains that utilize similar neural compensation mechanisms.

\section{Conclusion} \label{sec:conclusion}
In this work, we introduced PhyNC—a novel deep learning framework that seamlessly integrates a neural field with a physical forward model to achieve high-resolution conductivity reconstruction. Our unsupervised approach eliminates the dependency on external training datasets, enabling robust generalization from coarse to fine meshes while compensating for the uneven sensitivity intrinsic to physical models.

By leveraging a hybrid representation that combines Fourier features with Mip-Map embeddings, PhyNC ensures both stability and enhanced reconstruction fidelity. Extensive comparisons with existing baselines demonstrate its superior capability in capturing shape details and mitigating artifacts, even when working with coarse meshes. Moreover, our analysis of sensitivity intervals and the adaptive learnable embedding space underscores the importance of incorporating a structural prior, which significantly contributes to the model's overall performance.

Overall, PhyNC not only advances the state-of-the-art in tomographic imaging techniques like EIT but also opens avenues for future research aimed at further improving performance in regions with low sensitivity and extending the framework to other complex imaging modalities.
\bibliographystyle{unsrt}  

\bibliography{references}

\end{document}